\def\be{\begin{equation}}
\def\ee{\end{equation}}
\def\ba{\begin{eqnarray}}
\def\ea{\end{eqnarray}}
\documentstyle[preprint,aps,floats]{revtex}
\tightenlines
\draft  
\begin{document}
\title{Non-perturbative scaling in the scalar theory}
\author{Alfio Bonanno}
\address{Osservatorio Astrofisico, Universit\`a di Catania\\
Viale Andrea Doria 6, 95125 Catania, Italy\\
INFN, sezione di Catania\\
Corso Italia 57, 95128 Catania, Italy\\
}
\maketitle

\draft

\begin{abstract}
A new approach to study the scaling behavior of the scalar theory near the 
Gaussian fixed point in $d$-dimensions is presented. For a class of initial data 
an explicit use of the Green's function of the evolution equation is made.
It is thus discussed under which conditions non-polynomial relevant interactions  
can be generated by the renormalization group flow.
\end{abstract} 
\narrowtext
Recent studies have discussed the possibility that at the Gaussian Fixed Point (GFP) of
the $O(N)$ invariant scalar theory new relevant eigen-directions exist. 
In particular in \cite{huang1} Halpern and Huang have found a class of 
non-polynomial relevant interactions at the GFP by means of 
Wilson's renormalization group (RG) transformation with the ``sharp'' momentum
shell integration, but their result is still much debated \cite{morris1,periwal}.

In fact an interesting question to study is the general 
features of the scaling around the 
GFP. In particular it would be important to have a global understanding of all the possible
solutions in order to discuss the structure of the continuum limit. 

A necessary condition to describe this  limit 
is to select a set of marginal 
and relevant scaling eigen-operators at this fixed point.
A scaling eigen-operator is a solution of the linearised 
fixed point equation of the form 
\be\label{0}
u = \left({\Lambda \over k}\right )^\nu h(x)
\ee
where $x$ is the dimensionless field, $\Lambda$ is the UV overall cut-off,
$k$ is the running cut-off and $\nu$ is a scaling exponent. 
While in perturbation theory only a finite number of polynomial relevant 
and marginal interactions are considered, it is still not clear if new non-perturbative
non-polynomial relevant scaling interactions 
can be generated by the RG flow.
Were this true, it would mean for instance that the polynomial interactions
cannot span all the continuum physics.

The aim of this work is to clarify some aspects of this question. 
Instead of using the methods discussed in \cite{huang1,hasen,aboza}, 
the Green's function of the linearised renormalization
group equation is constructed. It will be thus shown the role
played by the boundary conditions in determining the structure of the 
scaling fields and it will be shown that  
is possible to generate relevant non-polynomial interactions 
not belonging to the same universality class of the polynomial interactions. 

The most convenient way to address such a question is to use the Wilsonian formulation of 
the RG transformation. 
In particular we consider the local potential approximation (LPA) of 
Wegner-Houghton \cite{wegner} equation as discussed in \cite{hasen}
for the  $N=1$ components scalar theory, but we shall not restrict the 
potential to a polynomial. 
 
The flow equation for the scaling field $u(x,t)$ reads \cite{hasen,aboza}  
\be\label{1}
{\partial u\over \partial t}=d \; u-{d-2\over 2}\; x \; 
{\partial u \over \partial x} +a_d \; {\partial ^2 u\over \partial x^2}  
\ee
where $d>2$ is the dimension of the spacetime,  $\Lambda$ is the UV cut-off,  
$a_d=1/({2 \sqrt{\pi}})^d\Gamma(d/2)$, and $t\equiv {\rm ln} \; {\Lambda / k}$.
The GFP is at $u\equiv 0$. 
In studying the UV region near the GFP one is interested in 
finding the unique function $u(x,t)$ which is 
continuous in the closed upper half plane $-\infty < x <\infty$, $0\leq t$  and satisfies
eq.(\ref{1}) with initial condition
$u(x,0)=f(x) $, $ -\infty <x<\infty$
with $f(x)\in C^0$. We shall leave open the possibility that 
$f$ is not bounded since there are no physical reasons to assume  
$|u(\pm \infty,t)|<\infty$ in discussing the scaling interactions. 

It is convenient to make the transformation
\be\label{2}
u(x,t) =v(x,t)\; e^{(3d-2)t/ 2}\\
\ee
which brings (\ref{1}), in the equivalent Fokker-Planck form \cite{risken}
\be\label{3}
{\partial v\over \partial t}=-{d-2\over 2} \; 
(v+x \;{\partial v\over \partial x}) +a_d\;v_{xx}\\[2mm]
\ee
After the rescaling $x \rightarrow x \; \sqrt{(d-2)/2 a_d}\;$ and 
$t\rightarrow t\; (d-2)/2\equiv \tau$, it reads
\ba\label{4}
{\partial v\over 
\partial \tau} = L_{\rm FP}[v],\hspace{2cm}
L_{\rm FP}\equiv 
{\partial \over \partial x} \left [ {\partial \over \partial x} -x\right ]
\ea
where we have introduced the Fokker-Planck operator $L_{\rm FP}$. 
It is not difficult to build a Green's function for such an operator. 
Let us define the formally self-adjont operator
\be\label{5}
L = e^{-{x^2/4}} L_{\rm FP} e^{{x^2/4}}=
-\left [-{\partial^2\over \partial x ^2} +{x^2\over 4}+{1\over 2} \right ]
\equiv -H_{\rm ho}-{1\over 2}
\ee
being $H_{\rm ho}$ the Hamiltonian of the quantum mechanical harmonic oscillator 
with $\hbar=\omega=1$ and mass $m=1/2$.  We shall refer to the description in
terms of such an operator and eigenfunctions as the quantum-mechanical 
(QM) frame, as opposed to the Fokker-Planck (FP) frame in eq.(\ref{4}).  The eigenfunctions
of $L$ have the asymptotic behavior $\sim e^{\pm x^2/4}$ as 
$|x|\rightarrow \infty$. For $L$ to be self-adjoint only the damped exponential 
must be chosen, and the eigenfunctions are know from Quantum Mechanics.
In particular the spectrum of $L$ satisfies $L\; \varphi_n = -(n+1)\varphi_n $
with $n=0,1,2,\cdots$ and all the eigenvalues of $L$ are eigenvalues 
of $L_{\rm FP}$ with eigenfunctions 
$L_{\rm FP} \phi_n = - (n+1)  \phi_n = e ^{x^2 / 4}\varphi_n$
Therefore a complete and orthonormal set of eigenfunctions for the $L_{\rm FP}$ 
operator reads  
\be\label{8}
\phi_n(x)\; = \; {1\over ^4\sqrt{2\pi}\;\sqrt{2^n n!}} \; H_n(x/\sqrt{2})
\ee
where $H_n$  are the Hermite polynomials, with the orthonormality condition  
\be\label{9}
\langle \phi_n|\phi_m\rangle
=\int_{-\infty}^{+\infty} e^{-x^2/2}\; \phi_n\phi_m \; dx \; = \;\delta_{mn}
\ee
Thus the linear space generated by the set $\phi_n$ is ${\cal L}^2({\cal R})$ with the 
norm $||f||=\sqrt{\langle f|f\rangle}$. 
It is now straightforward to write down the Green's function $G(x,\tau|y,0)$ of the 
Fokker-Planck operator from the well known expression of the Green's function 
of the harmonic oscillator. We have
\ba
&&G(x,\tau|y,0) = \langle x| e^{\tau L_{\rm FP}}| y \rangle = e^{-{y^2 / 2}}
\sum_n e^{-\tau(n+1)} \phi_n (x)\phi_n(y)\\[2mm]\label{fp1}
&&=e^{-{y^2/ 2}}e^{x^2/ 4}\sum_n e^{-\tau(n+1)} \varphi_n(x)\varphi_n(y) 
e^{y^2/ 4}\\[2mm]\label{fp2}
&&=e^{x^2/ 4} \langle x | e^{-\tau (H_{\rm ho}+{1\over 2}) }| y 
\rangle  e^{-{y^2/ 4}-\tau/2}
\\[2mm]\label{fp3}
&&=\sqrt{{e^{-\tau}\over 4\pi\;{\rm sinh \tau}}}
\exp\left (-{e^\tau\over 4 \;{\rm sinh \tau}}\; (x\;e^{-\tau}-y)^2\right )
\ea
This precisely what we find in \cite{risken} with a summation formula
of the Hermite polynomials.  Therefore
the unique solution of (\ref{4}) which is continuous and satisfies the following 
initial conditions $v(x,0) =f(x)$ with $|\langle \phi_n | f \rangle | < \infty$
is given by 
\be\label{10}
v(x,\tau) = \int_{-\infty}^{+\infty} G(x,\tau;y,0) f(y) \; dy
\ee

We now come to the main questions of our investigation, namely to study the 
subspace of relevant and marginal interactions present at the fixed point.
It is not difficult to see that 
{\it the solutions of eq.(\ref{1}) that $\in {\cal L}^2({\cal R})$
which are not bounded as $t \rightarrow +\infty$ have
$n\leq 2d/(d-2)$ in the spectrum of $L_{\rm FP}$.} 

The proof is a consequence of the spectral properties of the flow evolution operator:
given $f\in {\cal L}^2({\cal R})$ as an initial condition,
from (\ref{2}) and (\ref{10}) one has
\ba
&&u(x,t) = e^{(3d-2)t/2}\int^{+\infty}_{-\infty} G(x,\tau|y,0) f(y)\\[2mm]
&&=e^{(3d-2)t/2}\int_{-\infty}^{+\infty} 
\sum_n e^{-\tau(n+1)} \phi_n(x)\phi_n(y) f(y) 
e^{-{y^2\over 2}}dy\\[2mm]
&&=\sum_n e^{(d-{(d-2)n/2})t} \phi_n(x) \langle \phi_n | f \rangle
\ea
where we have used eq.(\ref{fp2}) and the uniform convergence of the sum. From 
the last line we see that in order to have a solution that grows for $t>0$  it must be
$n\leq {2 d/ (d-2)}$. If we set $f=\phi_m$ we recover the 
subspace of marginal and relevant eigenvectors spanned
by the first $n\leq 2d / (d-2)$ Hermite polynomials,  as it has already found in the 
previous literature \cite{hasen,aboza}. 
Non-polynomial interactions may be evolved by means of the Green's function 
provided $|\langle \phi_n | f\rangle |< \infty$,  but these will not 
be of the form (\ref{0}) and therefore cannot be considered as scaling fields.
More specifically if we try to evolve an initial datum that behaves like 
$\exp(qx^2)$ for $x\rightarrow \infty$ with $q<1/2$, we find
$u(x,t)\sim e^{(3d-2)t/2} \exp(e^{-(d-2)t} q x^2)$ as $t\rightarrow \infty$
(note that $|\langle \phi_n | f\rangle |=\infty $ if $q\geq 1/2$. ).

We want now to show that  the space where all the possible physically meaningful
scaling interactions live is greater than ${\cal L}^2({\cal R})$.
The key point to be noticed is that {\it the QM frame does not reproduce 
all the  possible solutions present in the FP frame}.  This is because 
$L$ is self-adjoint for  the boundary conditions at infinity of the type 
$\varphi_n\rightarrow e^{-x^2/4}$.  In this case from Sturm-Liouville 
theory we know that the spectrum is bounded and the number of relevant 
and marginal interactions is thus finite \cite{morris3}.  
However $L_{\rm FP}$ is not self-adjoint and its spectrum is greater than the 
self-adjoint extension of $L$. In fact, although there are no non-stationary solutions 
of the linearised equation in the QM frame, a non-trivial zero mode is 
present in the spectrum of the 
$L_{\rm FP}$ operator, namely
\be\label{sc1}
v_0 = e^{x^2/ 2}
\ee
which of course $\not\in {\cal L}^2({\cal R})$ and it corresponds 
to the following non-stationary solution of the original equation 
(\ref{1})
\be\label{sc2}
u(x,t) = e^{(3d-2)t/2}\;  \; e^{(d-2)x^2/2a_d} 
\ee
where we have inserted the factor $\sqrt{(d-2)/2a_d}$ in the definition of $x$.
This is a perfectly well defined and regular scaling field which is ``relevant" 
because it grows in the IR, with scaling dimension $\nu = (3d-2)/ 2$, and it does
not belong to the subspace spanned by the $\phi_n$ eigenfunctions previously 
constructed.  
It is therefore a non-perturbative scaling field (the connection of this solutions 
with the type discussed in \cite{huang1} is not immediately clear to me
although their asymptotic behavior for large $x$ is similar to what 
discussed in \cite{huang1}.).
In Quantum Mechanics or in Ornstein-Uhlenbeck diffusion
processes one would discard such a solution because of the boundary conditions at 
infinity. In our context there is no specific reason for not considering
scaling fields of this type.  

Given this particular zero mode of the $L_{\rm FP}$ operator it is immediate to 
write down the general solution of the homogeneous equation
$L_{\rm FP}[v] = 0$ which is of the form
\be
v(x,\tau) =e^{x^2/2}
\left( A+B \; {\rm erf}\;({x\over \sqrt{2}})\right ) 
\ee 
being $A$ and $B$ two arbitrary constants of integration. More generally if
we set $v_n(x,\tau)=e^{\lambda \tau }h_\lambda(x)$ eq.(\ref{4}) reads 
\be\label{homo}
h''(x)-xh'(x) -(\lambda+1)h(x) = 0
\ee
For integer and non-negative values of $\lambda$ a particular solution of the homogeneous 
equation (\ref{homo}) can always be found with the Laplace method.
One explicitly finds for $\lambda=1,2,3,4$
\ba\label{scaling}
&&h_1=e^{x^2/2}\; x   \;\;\;\;\;\;\;\;\;\;\;\;\;\;\;\; h_2 = e^{x^2/2}(1+x^2) \\[2mm]
&&h_3 = e^{x^2/2}(3x+x^3) \;\;\;\;\; h_4 = e^{x^2/2}(3+6x^2+x^4)\nonumber 
\ea
It should be stressed that solutions of the type (\ref{scaling}) which are even function of $x$
are bounded from below and a linear combination may develop non-trivial minima at 
some value of the field. This is the case of $h_4 e^{4\tau}+c\;h_2e^{2\tau}$ 
for some negative values of the constant $c$. The general solution is 
\be
v_n(x,\tau) = e^{n\tau}
\left (A \; h_n(x)+B \; h_n(x)\int^x e^{s^2/2}\; h_n(s)^{-2} ds\right )
\ee
where $h_n$ is obtained with the Laplace method and $n=0,1,2,\cdots$. 

The presence of solutions of the flow equation with entirely different physical implications
should not come as a surprise. In $d=2$ eq.(\ref{1}) admits the following
solutions, as it can be checked by direct substitution,
\ba
&&u(x,t) = e^{(2-\beta^2/4\pi)t}\cos(\beta x)\\[2mm]
&&u(x,t) = e^{(2+\gamma/4\pi)t}e^{\sqrt{\gamma} x}
\ea
where $\beta$ and $\gamma$ are arbitrary constants.
The first one describes the physics of the Sine-Gordon
theory, while the second one defines the Liouville theory
\cite{reuwe}. Also in this case the same fixed point may describe 
entirely different physical theories. 

The physical relevance (if any) of these solutions is not clear to us. Although 
one may speculate about their possible applications to the Standard Model or 
to Cosmology, all we can say for the moment is that the structure of the GFP 
for a simple scalar theory is richer than previously discussed. 
In particular, the universality class spanned by the polynomial eigen-potentials
in the LPA does not contain all the possible relevant interactions. 

One final remark is the following.  We have used the linearised Wegner-Hougton
equation in the LPA approximation with sharp cut-off. This equation 
is the same of the Polchinski equation obtained with the smooth cut-off, apart
for the unimportant constant $a_d$ \cite{periwal}.  
Thus our result is quite robust  since these new scaling 
scaling exponents does not depend on the cut-off function used in the derivation. 

\vspace {1cm}
I am very much indebted with Martin Reuter and Dario Zappal\`a for very enlightening
discussions and encouragements.

\end{document}